\documentclass{article}

\usepackage[english]{babel}

\usepackage[letterpaper,top=2cm,bottom=2cm,left=3cm,right=3cm,marginparwidth=1.75cm]{geometry}

\usepackage{amsmath}
\usepackage{graphicx}
\usepackage[colorlinks=true, allcolors=blue]{hyperref}

\title{Impact of the Consistency Model on \\ Checkpointing of Distributed Shared Memory\thanks{This work is funded in part by award 1849599 and award CNS1816487 from the National Science Foundation. The views and conclusions contained in this document are those of the authors and should not be interpreted as representing the official policies, either expressed or implied, of the National Science Foundation or the U.S. Government.}}
\author{Sandeep Kulkarni, Duong Nguyen, Lewis Tseng, Nitin Vaidya\thanks{The authors are listed alphabetically. Sandeep Kulkarni is at the Michigan State University, Lewis Tseng is at Boston College, and Nitin Vaidya is at Georgetown University. This work was done when Duong Nguyen worked at Georgetown University.}}

\begin{document}
\maketitle

\begin{abstract}
In this report, we consider the impact of the consistency model on checkpointing and rollback algorithms for distributed shared memory. In particular, we consider specific implementations of four consistency models for distributed shared memory, namely, linearizability, sequential consistency, causal consistency and eventual consistency, and develop checkpointing and rollback algorithms that can be integrated into the implementations of the consistency models. Our results empirically demonstrate that the mechanisms used to implement stronger consistency models lead to simpler or more efficient checkpointing algorithms.
\end{abstract}

\section{Distributed Shared Memory}

In this report, we consider the following architecture for distributed shared memory. The system consists of $n$ nodes. Each node consists of a replica of the shared memory as well as a client. A client only communicates with the local replica on the same node. The replicas communicate with each other to implement the desired shared memory consistency model. In the following sections, we will consider specific implementations of the following four consistency models, and present checkpointing and rollback algorithms that can be integrated into these implementations.
\begin{itemize}
    \item Linearizability \cite{DBLP:books/daglib/0017536}
    \item Sequential consistency \cite{DBLP:books/daglib/0017536}
    \item Causal+ consistency \cite{DBLP:conf/sosp/LloydFKA11}
    \item Eventual consistency \cite{DBLP:journals/cacm/Terry13}
\end{itemize}

We consider fully replicated implementations of distributed shared memory. Thus, each replica of the shared memory maintains its own copy of each shared variable.

The clients can invoke two shared memory operations, namely write(variable, value) and read(variable). For instance write(X,3) is performed when the client wants to write 3 to shared memory variable X. read(X) is performed when a client wants to read shared memory variable X. A client invokes an operation by sending a message to its local replica, specifying the desired operation. When the client receives a response (an ``{\em Ack}'') from the local replica, the invoked operation is said to be completed. The client communicates with its local replica over FIFO channels. When a client delivers a message from the replica, it completes all necessary steps in response to the message before delivering another message from the replica.
 
Each client has local (unshared) state, and a ``program counter". The local state of a client also includes the most recent response received from the local replica. When the client has invoked an operation that is still incomplete, the program counter is the index of the incomplete operation. Otherwise, the program counter is the index of the next operation to be invoked by the client. A client may have at most one incomplete operation at any given time. We will use the following terminology:
\begin{itemize}
    \item State of replica $i$: State of the copies of shared memory variable stored in replica $i$. 
    \item Local state of client $i$: State of unshared variables at client $i$.
\end{itemize}

The presentation of the implementation of linearizability and sequential consistency models below borrow text from \cite{vaidya-notes}, which, in turn, base the presentation on the treatment in Attiya and Welch \cite{DBLP:books/daglib/0017536}.

\paragraph{Correctness:}
For checkpointing and rollback, we need to define what constitutes ``correctness''. Intuitively, we would want the execution to remain faithful to the desired
consistency model, despite rollbacks.

From each client's perspective, our rollback procedures will ensure that the Read/Write operations have ``exactly once'' semantics -- that is, to the client, each operation it invokes will appear to have been performed exactly once, and in the program order (even if rollbacks occur). Then, for correctness, we require that the resulting execution (i.e., Read/Write operations performed by all the clients) satisfies the desired consistency model.

\paragraph{Past work:} While there is much work on checkpointing of distributed shared memory, we are not aware of work that takes into account the impact of the consistency model implementation on the checkpointing/rollback algorithms. \\~\\

In the rest of this report, we consider each of the four consistency models above.
\section{Linearizability}

\subsection{Implementation of Linearizability}

We now present an implementation of linearizability, which is based on the treatment in Attiya and Welch \cite{DBLP:books/daglib/0017536}. We assume that a totally-ordered multicast primitive is available to the replicas, which also provides the FIFO property. Thus, all the replicas deliver all totally-ordered multicast messages in the same order, and the totally-ordered multicasts invoked by any single replica are delivered in the order of their invocation.

\paragraph{When client $i$ invokes operation {\tt Write(X,v)}:}
\begin{list}{}{}
\setlength\itemsep{0pt}
\item[(step w1)] Replica $i$ performs a {\em totally-ordered multicast} of $Write(X,v)$.
\item[(step w2)] When its own totally-order multicast message $Write(X,v)$ is delivered to replica $i$, the replica writes $v$ to its copy of shared variable $X$.
\item[(step w3)] Send {\em Ack()} to client $i$.
\end{list}
\paragraph{When replica $i$ delivers totally-ordered multicast message $Write(Y,w)$ for
multicast performed by some other replica $j$ ($j\neq i$):}
\begin{list}{}{}
\item[(step w4)] Write value $w$ to the local copy of shared variable $Y$ at replica $i$.
\end{list}
\paragraph{When client $i$ invokes operation {\tt Read(X)}:}
\begin{list}{}{}
\item[(step r1)] Replica $i$ performs a {\em totally-ordered multicast} of $Read(X)$.
\item[(step r2)] When totally-ordered multicast of $Read(X)$ initiated by replica $i$ is delivered to replica $i$,
read the value of local copy of X at replica $i$; suppose that the value read is $v$.
\item[(step r3)] Send {\em Ack(X,v)} to client $i$ as the response to {\it Read(X)}.
\end{list} 
\paragraph{When replica $i$ is delivered totally-ordered multicast of $Read(Y)$ initiated by some
other replica $j$ ($j\neq i$):}
\begin{list}{}{}
\item[(step r4)] No action is needed. Discard the message.
\end{list}

When a replica delivers a totally-ordered multicast message, it completes all steps required to handle that message before handling to any subsequent totally-ordered multicast message that may be delivered. In particular, it must complete steps w2-w3 or w4 or r2-r3, depending on the type of message delivered, before delivering another message.

\subsection{Checkpointing for Linearizability}

We will assume that one of the replicas is designated as the {\em initiator} (denoted I) of distributed checkpointing algorithm. The goal of the checkpointing algorithm is to record the state of the system (i.e., clients, replicas and the communication network, if necessary) such that a consistent state can be restored after a failure (``rollback''). The checkpointing algorithm can exploit the specific communication structure among the replicas, as used by the consistency algorithm. The use of {\tt Marker} messages below is motivated by analogous messages used in the Chandy-Lamport snapshot algorithm \cite{DBLP:journals/tocs/ChandyL85}.

\paragraph{Initiator's steps:} The initiator I initiates the checkpointing algorithm by performing a totally-ordered multicast of a {\tt Marker} message to all the replicas. This message is totally-ordered with respect to the Read and Write messages described above. When replica I delivers the {\tt Marker} message, it atomically performs the following steps:
\begin{itemize}
    \item (Replica Step 1) The replica records its own state, i.e., replica state.
    \item (Replica Step 2) The replica sends a {\tt Marker} message to its local client, and awaits an {\em Ack} message from the client. On receiving a {\tt Marker}, the client then records its own local state, and its program counter as follows.
    \begin{itemize}
        \item If the client has invoked an operation that does not access shared memory, the client first completes that operation. The recorded program counter corresponds to the next operation to be performed by the client.
        \item If the client has invoked a shared memory operation for which an {\em Ack} is not received yet, the program counter will correspond to this incomplete operation.
        \item If the client has not invoked any operation at the time it receives the {\tt Marker}, then the recorded program counter corresponds to the next operation to be invoked by the client.
        \end{itemize}
        When the above steps are complete the client sends an {\em Ack} to the replica.
    \end{itemize}

\paragraph{Steps at any other replica $j$ ($j\neq$ I):} When replica $j$ delivers the {\tt Marker} message, it performs Replica Step 2 described above. \\

\subsection{Rollback for Linearizability}

After a failure is detected, the state of the system is rolled back to the state recorded during the most recent successfully completed checkpoint operation. In particular:
\begin{itemize}
    \item The multicast layer state is cleared. In other words, after rollback, there will not be any multicast operations in progress, and no buffered multicast messages anywhere in the system. Thus, any multicast messages delivered after rollback will necessarily be the result of the re-execution that occurs after the rollback.
    \item The state of each replica is restored to the state recorded by the initiator I during the checkpoint.
    \item The local state and program counter of any client $j$ are restored to the local state and program counter recorded during the checkpoint.
\end{itemize}
The above steps are performed during rollback for all the consistency models discussed in this report, and we will not repeat these steps elsewhere below.

Note that the program counter of a client $j$ may correspond to a shared memory operation that was invoked but not completed at the time the checkpoint was taken. This operation will be invoked yet again after the rollback is completed. 
\begin{itemize}
    \item Incomplete Read operation: Since the Read operation was incomplete when the checkpoint was taken, the Read did not modify the local state of the client when incompletely performed. 
    Thus, re-issuing the Read after rollback is appropriate.
    \item Incomplete Write operation: Since the Write was incomplete, the client has not yet received an {\em Ack} message from the replica.
    This, in turn, implies that the local replica did not deliver the corresponding totally-ordered Write message before delivering its own {\tt Marker} message -- in fact, none of the replicas delivered the Write message before delivering the {\tt Marker} message from the initiator. Thus, the replica states were not modified by the incomplete Write operation. Thus, it is appropriate to re-issue the incomplete Write operation.
\end{itemize}

\paragraph{Correctness:} Observe that, due to a rollback, the effect of any client $j$ operations that are completed after its local replica $j$ has recorded its state during checkpointing, but before rollback occurs, is undone. That is, any changes to the local state of the client due to such operations is undone, and changes to the state of any replicas due to such operations is undone. 

Operations of client $j$ that were completed before replica $j$ recorded its state are not re-issued after rollback. Thus, each operation by any client receives only one response, and ``exactly once'' semantics is achieved for the client operations. Also, the operations also appear to have been performed in program order (as noted above, the effect of any operations that complete between checkpoint and rollback is undone).

Since the operations appear to have been completed in the total order of the delivery of the corresponding multicast messages (excluding multicasts for any operations that are undone), the resulting execution will satisfy linearizability.

\subsection{Discussion}

By exploiting the fact that all Read and Write operations are totally-ordered, we are able to checkpoint the shared memory by requiring just one replica (the initiator) to record the replica state. A direct implementation of Chandy-Lamport snapshot algorithm will require each replica state to be recorded -- this is unnecessary, because all replicas have an identical state at the point when the {\tt Marker} message is delivered to the replica. This illustrates that the chosen implementation of the stronger consistency model allows a more efficient checkpointing algorithm. It remains to be shown that such a benefit is obtained for {\em any} implementation of the stronger consistency model.

\section{Sequential Consistency}

\subsection{Implementation}

The implementation below is based on the treatment in Attiya and Welch \cite{DBLP:books/daglib/0017536}.

\paragraph{When client $i$ invokes operation {\tt Write(X,v)}:}
\begin{list}{}{}
\setlength\itemsep{0pt}
\item[(step w5)] Replica $i$ performs a {\em totally-ordered multicast} of $Write(X,v)$.
\item[(step w6)] When its own totally-ordered multicast message $Write(X,v)$ is delivered to replica $i$, it writes value $v$ to the local copy of $X$ at $p_i$.
\item[(step w7)] Send {\em Ack()} to client $i$.
\end{list}
\paragraph{When replica $i$ delivers totally-ordered multicast message $Write(Y,w)$ for
multicast performed by some other replica $j$ ($j\neq i$):}
\begin{list}{}{}
\item[(step w8)] Write value $w$ to the local copy of $Y$ at $p_i$.
\end{list}
\paragraph{When client $i$ invokes operation {\tt Read(X)}:}
\begin{list}{}{}
\item[(step r5)] Read local copy of variable $X$ at replica $i$; suppose that the value read is $v$.
\item[(step r6)] Send {\em Ack(X,v)} to client $i$ as the response to {\tt Read(X)}.
\end{list} 

When a replica is delivers a {\em Write} totally-ordered multicast message, it completes either steps w6-w7 or step w8, as appropriate, before delivering another message.

\subsection{Checkpointing for Sequential Consistency}

The checkpointing procedure is identical to that described for linearizability. In this case, the {\tt Marker} messages
are totally-ordered with respect to only the Write messages (since there are no totally-ordered Read messages for sequential consistency, unlike linearizability).

\subsection{Rollback for Sequential Consistency}

The rollback procedure here is identical to that for linearizability. Correctness argument is analogous to that for sequential consistency.

\subsection{Discussion}

Similar to linearizability, for sequential consistency as well, only the initiator needs to record the replica state. Sequential consistency is a weaker model than linearizability. So, a natural question is whether there any gain in efficiency of checkpointing with sequential consistency, when compared to linearizability. Of course, it is more efficient to implement sequential consistency. However, there does not seem to be any nontrivial improvement in the cost of checkpointing with sequential consistency.

\section{Causal+ Consistency}

We consider an implementation of Causal+ consistency model proposed in the work on COPS \cite{DBLP:conf/sosp/LloydFKA11}. The Causal+ consistency
model is stronger than causal consistency, and also provides the property of convergence, i.e., all the replicas 
have the same values for the shared memory variables after updating an identical set of updates, even if the updates are not
applied in an identical order.

\subsection{Implementation}
We assume that the replicas maintain a vector timestamp, which records the number of Writes (i.e., updates) from each replica that have been applied 
up to any given time. Thus, if there are 4 replicas, and replica 1 has so far applied 3, 2, 1 and 5 updates from replicas 1, 2, 3 and 4,
respectively, then the vector timestamp of replica 1 will be [3,2,1,5]. Such vector timestamps were introduced in prior work
on shared memory \cite{DBLP:journals/tocs/LadinLSG92}. In addition, we assume that the replicas also maintain Lamport timestamps. The tuple (Lamport timestamp,replica id) 
is used to establish a total order on the updates issued by the different replicas -- this total order will help achieve
the convergence property (this approach is also used in COPS \cite{DBLP:conf/sosp/LloydFKA11}).
\begin{itemize}
\item It is implicitly assumed that all the messages are piggybacked
with the appropriate vector and Lamport timestamps.
\item In each replica, each shared memory variable is tagged with the vector timestamp and Lamport timestamp of the update
message that resulted in the update of that variable.
\end{itemize}
Clearly, the above approach is expensive (in particular, tagging each variable with the timestamps requires significant storage).
It is possible to make the implementation of the consistency protocol more efficient, while implementing a checkpointing
scheme similar to that described below. For ease of presentation, we use the less efficient approach. 

\paragraph{When client $i$ invokes operation {\tt Write(X,v)}:}
\begin{list}{}{}
\setlength\itemsep{0pt}
\item[(step w9)] Replica $i$ performs a {\em causally-ordered multicast} of $Write(X,v)$.
\item[(step w10)] When its own causally-ordered multicast message $Write(X,v)$ is delivered to replica $i$, it writes value $v$ to the local copy of $X$ at $p_i$, provided that the (Lamport timestamp, $i$) tuple tagged to the message is larger than the corresponding tag for $X$ at replica $i$.
\item[(step w11)] Return {\em Ack()} to client $i$ indicating completion of the {\tt Write(X,v)} operation 
\end{list}
\paragraph{When replica $i$ delivers causally-ordered multicast message $Write(Y,w)$ for
multicast performed by some other replica $j$ ($j\neq i$):}
\begin{list}{}{}
\item[(step w12)] Write value $w$ to the local copy of $Y$ at $p_i$  provided that the (Lamport timestamp, $j$) tuple tagged to the message is larger than the corresponding tag for $Y$ at replica $i$.
\end{list}
\paragraph{When client $i$ invokes operation {\tt Read(X)}:}
\begin{list}{}{}
\item[(step r7)] Read local copy of variable $X$ at replica $i$; suppose that the value read is $v$.
\item[(step r8)] Return {\em Ack(X,v)} to client $i$ as the response to {\tt Read(X)}, which also indicates the
completion of the {\tt Read(X)} operation.
\end{list}

\subsection{Checkpointing for Causal+ Consistency}

\paragraph{Initiator's steps:} The initiator I initiates the checkpointing algorithm by performing a causally-ordered multicast of  {\tt Marker}($V_I$) message, where $V_I$ is the vector timestamp at the initiator. The $j$-th element, $V_I[j]$, is the number of writes by replica $j$ (or more precisely, by client $j$) that have been performed at replica $I$ so far.
This {\tt Marker} message is causally-ordered with respect to the Write messages described above. When replica I delivers its own {\tt Marker} message, it atomically performs Replica Steps 1 and 2 described for linearizability. The client behavior is identical to that in the checkpoint procedure for linearizability.

\paragraph{Steps at any other replica $j$ ($j\neq$ I):}

Let $V_j$ denote the vector clock at replica $j$ when it delivers a {\tt Marker}($V_I$) message from replica I. Then the following inequality will hold due to the causally-ordered broadcast used for the {\tt Marker}: $V_j[k] \geq V_I[k]$, for $k\neq I$ and $V_I[I]=V_j[I]+1$.

When {\tt Marker}($V_I$) is delivered to replica $j$, it performs the following steps
atomically. It records own updates (Writes) with index greater than $V_I[j]$ (i.e., these updates were not propagated to replica $I$ by the time replica $I$ sent
the {\tt Marker} message). In effect, replica $j$ fills the gap between the updates it has issued and the updates replica $I$ has received from $j$.
Replica $I$ then performs Replica Step 2 described earlier for linearizability.


\subsection{Rollback for Causal+ Consistency}

Clients roll back to their recorded state. Each replica is first initialized to the state recorded by the initiator I. Then, the incremental updates recorded by each replica $j\neq I$ are sequentially applied at each replica, provided that this is permitted by the convergence procedure (using Lamport timestamps). At the end of these steps, all replicas will be in an identical state. The use of Lamport timestamp will also prevent a replica from applying an update that it has already applied before it recorded own state during the checkpoint algorithm.

\paragraph{Correctness:} 
The exactly once semantics and program order of client operations despite rollback is similar to linearizability. However, rollback for Causal+ is somewhat different than sequential consistency and linearizability in that, after rollback, a given replica $j$ may end up applying more updates from another replica $k$ as compared to the number of replica $k$'s updates that replica $j$ had applied when it recorded its state during the checkpoint. So, we need to argue that the resulting execution still satisfies Causal+ consistency model.

We argue this inductively. Suppose that the execution that takes place before a failure occurs (resulting in a rollback) is Causal+ consistent. Let us denote by $V_i$ the vector timestamp of any replica $i$ when it recorded its own state during the most recently completed checkpointing (prior to the failure). Because Write messages are delivered using causally-ordered broadcast, it must be the case that $V_j[j]\geq V_i[j]$, $\forall i,j$.
During rollback, any replica $k$ will first restore its local state to the recorded state (i.e., its state with vector timestamp $V_k$), and then apply $V_j[j]-V_k[j]$ subsequent updates from each replica $j$. We make two observations: (i) All replicas end rollback after applying $V_j[j]$ updates from each replica $j$. Thus, due to the convergence property of the updates, all the replicas will complete rollback in the same state. (ii) At replica $k$, the last $V_j[j]-V_k[j]$ updates applied from replica $j$ must have been due to Writes by client $j$ that were concurrent with the next operation to be issued by replica $k$ after the rollback is complete (i.e., the operation pointed to by the program counter restored at client $k$ during rollback). This property holds due to the inductive assumption that the execution was Causal+ consistent prior to the failure. Then, applying the $V_j[j]-V_k[j]$ updates from replica $j$ before issuing any operation of client $k$ after rollback does not violate Causal+.

\subsection{Discussion}

Observe that replica $j\neq I$ needs to record some of its replica state, which is not required in sequential consistency and linearizability. Thus, the weaker consistency model results in higher checkpointing cost.

The convergence property of Causal+ lowers the cost of checkpointing by allowing any replica $j\neq I$ to only record its own updates that are not recorded in initiator's state. This optimization will not be available for causal consistency when the convergence property is not provided. This is mainly because that without the convergence property, replicas apply the same set of updates might not reach the same state, if they apply them in a different order.  Thus, the weaker causal consistency model will result in an increase in checkpoint cost, illustrating the trade-off alluded to earlier.

\section{Eventual Consistency}

Eventual consistency may be defined in a variety of ways. We will consider the version of eventually consistency that only requires the {\em FIFO} and {\em convergence} property: (i) updates from a given replicas will be applied in the order they are issued, and (ii) all replicas will eventually have an identical state after applying an identical set of updates.

\subsection{Implementation}

The implementation of eventual consistency is essentially identical to that for Causal+, with the only difference being that causally-ordered multicast is replaced by a FIFO-ordered multicast (which delivers messages from each replica in a FIFO order).

\subsection{Checkpointing for Eventual Consistency}

For eventual consistency, the checkpointing and rollback algorithms are similar to Causal+ consistency, with the only difference being in the transmission of {\tt Marker} messages. In fact, the checkpoint algorithm is substantially identical to Chandy-Lamport snapshot algorithm.

\paragraph{Initiator's steps:} The initiator I performs the following steps atomically. It initiates the checkpointing algorithm by sending  {\tt Marker}($V_I$) message to all the other replicas, where $V_I$ is the vector timestamp at the initiator. The $j$-th element, $V_I[j]$, is the number of writes by replica $j$ (or more precisely, by client $j$) that have been performed at replica $I$ so far.
The initiator then performs Replica Steps 1 and 2 described for linearizability. The client behavior is identical to that in the checkpoint procedure for linearizability.

\paragraph{Steps at any other replica $j$ ($j\neq$ I):} When {\tt Marker}($V_J$) is delivered to replica $j$ for the first time, it performs the following steps
atomically. It forwards {\tt Marker}($V_J$) to all the other replicas. It records own updates (Writes) with index greater than $V_j[I]$ (i.e., these updates were not propagated to replica $I$ by the time replica $I$ sent
the {\tt Marker} message).
Replica $I$ then performs Replica Step 2 described earlier for linearizability.

\subsection{Rollback for Eventual Consistency}

The rollback procedure is identical to that for Causal+ consistency.

\section{Discussion and Summary}

In this report, we have considered specific implementation for several different consistency models (Linearizability, Sequential Consistency, Causal+ consistency and Eventual Consistency), and presented checkpointing and rollback algorithms for these implementations. Theses algorithms empirically illustrate that weakening of the consistency model may increase the cost for the checkpointing algorithm.

The checkpointing algorithms for linearizability and sequential consistency save the state of just one shared memory replica. It can be argued that this is the minimal overhead required for any consistency model. We believe -- and hope to argue formally -- that the other checkpointing algorithms also save a minimal amount of replica state.


\end{document}